\documentstyle[prb,aps]{revtex}
\def\beq{\begin{eqnarray}}
\def\eeq{\end{eqnarray}}
\begin{document}
\draft

\title{THE FLUCTUATION INDUCED PSEUDOGAP IN THE INFRARED OPTICAL\\ CONDUCTIVITY 
OF HIGH TEMPERATURE SUPERCONDUCTORS}
\author{Francesca Federici and Andrei A. Varlamov \cite{note1}\\}
\address{Laboratorio "Forum" dell`Istituto Nazionale di Fisica della Materia,\\ 
Dipartimento di Fisica,
Universit\`a di Firenze,\\ Largo E.Fermi, 2,  50125 Firenze, Italy\\}
\date{\today}
\maketitle
\begin{abstract}
We study the effect of fluctuations on the {\bf ac} conductivity of a layered
superconductor both for $c$-axis and $ab$-plane electromagnetic 
wave polarizations. The fluctuation contributions of different physical
nature and signs (paraconductivity, Maki-Thompson anomalous contribution,
one-electron density of states renormalization) are found to be suppressed 
by the external field at different characterisitic frequencies 
($ \omega_{\rm AL}\sim T-T_{\rm c}$, $\omega_{\rm MT} \sim \max\{ T-T_{\rm c}, 
\tau_{\varphi}^{-1}\}$, $ \omega_{\rm DOS} 
\sim \min\{T,\tau ^{-1}\}$ for the $2D$ case). As a result the appearance of the
nonmonotonic frequency dependence (pseudogap) in the infrared
optical conductivity of HTS film is predicted. The effect has to be 
especially pronounced in the case of the electromagnetic field polarization 
along $c$-axis.

\end{abstract}

\section{Introduction}

Recently it was demonstrated \cite{ILVY,Gray,BMMVY} that due to such specific
properties of HTS as the extremely strong anisotropy and the
small coherence length the fluctuation renormalization of the one-electron 
density of states (or, in other words, the opening of the fluctuation 
pseudogap above $T_{\rm c}$) \cite{Abrah,DiCastro} can play an important role in 
the explanation of their normal phase properties near the superconducting 
transition. The depression of the one-electron density of states (DOS) on 
the Fermi level results in the decrease of related thermodynamic and 
transport properties which are proportional to that one. It is important 
that this effect has an opposite sign with respect to the traditionally 
discussed Aslamazov-Larkin (AL) and Maki-Thompson (MT) corrections though 
it remains less singular in $T-T_{\rm c}$ \cite{ILVY}. 

The most explicit manifestation of the effect discussed can be observed
in the $c$-axis transport properties of layered materials above $T_{\rm c}$, 
such as
the fluctuation conductivity and the magnetoresistance, where paraconductivity
and Maki-Thompson contribution in the $2D$ regime turn out to be suppressed
by the square of the interlayer trasmittance \cite{ILVY,BMV1,BDKLV,BMV2}. 
Another interesting example of the evident appearance of the negative DOS 
contribution is the case of the NMR relaxation rate for which the 
AL-like contribution is rigorously forbidden for the singlet pairing and the 
MT contribution compets with the DOS one \cite{AM,KF,RanVar}. 
Strong pairbreaking weaks the former and the latter leads to the decrease of 
the NMR relaxation rate at the edge of the transition \cite{Riga}.

In the present communication we attract the attention to another possibility 
to suppress the main AL and MT contributions by means of 
an high frequency electromagnetic field applied. Basing on the Drude origin 
of the normal part of the optical conductivity one can expect that noticeable 
changes in the electromagnetic wave reflectivity versus frequency take place 
for $\omega\sim\tau^{-1}$. But it turns out that the frequency dependence 
of paraconductivity and Maki-Thompson anomalous contribution in the vicinity 
of $T_{\rm c}$ appears long before, 
for $\omega \sim \min \{ T-T_{\rm c},\tau_{\varphi }^{-1}\}$ \cite{AV80,Sch}. 
These positive contributions compet with the 
negative $\sigma ^{\rm DOS}(\omega )$ which, as we will see below, weakly depends 
on frequency (only in the scale $\omega \sim \min\{ T,\tau ^{-1}\} $).
This competition results in the rapid decay of the dissipative processes 
already at frequencies of the order of $\omega \sim T-T_{\rm c}$ and  in the 
appearance of the transparency window up to $\omega\sim\min\{T,\tau^{-1}\}$. 
The following high frequency behavior of
 ${\rm Re}\left[\sigma (\omega )\right]$ is mostly
governed by $\sigma^{\rm n}(\omega )$ which decreases, in agreement  with 
Drude law, for $\omega\gtrsim\tau^{-1}$.

Below we will study the {\bf ac} fluctuation conductivity tensor for a 
layered superconductor taking into account all contributions and 
paying attention to the most interesting case of $c$-axis polarization of
the field. Nevertheless, for completness, the $ab$-plane results of the old 
paper of Aslamazov and Varlamov \cite{AV80} will be re-examined and 
discussed in application to the novel HTS layered systems.

\section{The Model and Some Definitions}

We study the optical conductivity of 
layered superconductors which can be expressed by the retarded analitical 
continuation of the current-current correlator (electromagnetic response 
operator) $Q^{(R)}(\omega)$ :
\beq
{\rm Re} \left[\sigma_{\alpha \beta}(\omega)\right] = - \frac{{\rm Im} 
\left[Q^{(R)}_{\alpha \beta}(\omega)\right]}{\omega} \label{Resig}
\eeq

The problem of fluctuation conductivity of layered superconductors has been
extensively studied and for the detailed description of the model and the
diagrammatic presentation of the electromagnetic response operator tensor
$Q_{\alpha\beta}$ we refer to \cite{BDKLV}. So below we will list only the 
necessary definitions and results following the notations of that article.

We assume the electron spectrum of the layered metal in the form  
\beq
\xi({\bf p}) =\epsilon_0({\bf p})+J\cos(p_zs)-E_{\rm F},\label{corrug}
\eeq
where $\epsilon_0({\bf p})={\bf p}^2/(2m)$, $p\equiv({\bf p}, p_z)$, 
${\bf p}\equiv(p_x,p_y)$ is a two-dimensional, intralayer wavevector, and $J$ 
is an effective quasiparticle nearest-neighbour interlayer hopping energy. 
The Fermi surface defined by the condition $\xi(p_{\rm F})=0$ is a corrugated 
cylinder, and $E_{\rm F}$ is the Fermi energy. This model spectrum is most 
appropriate for highly anisotropic layered materials, for which $J/E_{\rm F}\ll 1$. 
 
As it is well established now, the electron mean free 
path in HTS layered single crystals or epitaxial films turns out to be
of the same order or several times larger than the coherence length 
$\xi_{ab}(0)$, so the parameter $T_{\rm c}\tau \sim 1$ and the theory has to be 
constructed for an arbitrary impurity concentration. 

Intralayer quasiparticle scattering is included in the single 
quasiparticle normal state Green's function by means of the relaxation time
$\tau$:
\beq
G({\bf p},\omega_n)=\frac{1}{i\tilde\omega_n-\xi({\bf p})},\label{G}
\eeq
where $\tilde\omega_n=\omega_n[1+1/(2|\omega_n|\tau)]$.  

The "Cooperon" \cite{Alt} (triangle vertex accounting the interference in the 
impurity avaraging of the pairs of Green's functions), which is necessary in the 
following calculations, may be expressed, for an arbitrary impurity 
concentration, \cite{AV80,BDKLV} as:
\beq
C({\bf q},\omega_n,\omega_{n'})=\left[ 1-\frac{\Theta(-\omega_n\omega_{n'})}
{\tau(\tilde\omega_n-\tilde\omega_{n'})}\left(1-\frac{
\langle[\xi({\bf p})-\xi({\bf q}-{\bf p})]^2\rangle}
{(\tilde\omega_n-\tilde\omega_{n'})^2}\right)\right]^{-1}\label{C}
\eeq
where $\Theta(x)$ is the Heaviside step function, and $\langle\cdots\rangle$ 
denotes an average over the Fermi surface. Performing the Fermi surface 
average with the spectrum (\ref{corrug}), we find:
\beq
\langle[\xi({\bf p})-\xi({\bf q}-{\bf p})]^2\rangle=\frac12\left(v_{\rm F}^2{
\bf q}^2+4J^2\sin^2(q_zs/2)\right)\equiv \tau^{-1}D\hat{Q}^2 \label{D}
\eeq
where $v_{\rm F}=|{\bf p}_{\rm F}|/m$ is the magnitude of the Fermi velocity 
parallel to the layers.   In (\ref{C}), we have made the assumption
$\tau D\hat{Q}^2\ll 1$, which  we suppose throughout this manuscript and its 
limits have been discussed in \cite{RanVar}. The fluctuation propagator 
$L({\bf q},\omega_{\mu})$ in the vicinity of $T_{\rm c}$ has the form :
\beq
L^{-1}({\bf q},\omega_{\mu})=-\rho\left[\epsilon+\psi\left(\frac12+\frac{\omega_%
{\mu}}{4\pi T}+\frac{4\eta D\hat{Q}^2}{\pi^2v_{\rm F}^2\tau} \right)-\psi\left(
\frac12\right)\right], \label{L}
\eeq
where $\epsilon=\ln(T/T_{\rm c})\approx(T-T_{\rm c})/T_{\rm c}$ for $T-T_{\rm c}\ll T_{\rm c}$, $\psi(x)$ is 
the digamma function, $\rho=N(0)=m/(2\pi s)$ is the single-spin quasiparticle 
normal density of states, and
\beq
\eta =&-&\frac{v_{\rm F}^2\tau^2}{2}\left[\psi\left( \frac12+\frac{1}{4\pi\tau T}
\right)-\psi\left(\frac12\right)-\frac{1}{4\pi\tau T}\psi^{'}\left(\frac12
\right)\right] \label{eta} 
\eeq
is the positive constant which enters into the current expression in the
phenomenological {\it GL} theory in two dimensions \cite{GOR}.

\section{Paraconductivity}

Let us first examine the AL contribution (diagram 1 of Fig. 
\ref{FedericiPRB.1}) to the {\bf ac} fluctuation conductivity. 
The general expression for the 
appropriate contribution to the electromagnetic response operator as a 
function of the Matsubara frequencies of the external electromagnetic field 
$\omega_\nu$ is \cite{BDKLV}:
\beq \label{QAL}
Q^{\rm AL}_{\alpha\beta}(\omega_{\nu})&=&2e^2T\sum_{\omega_{\mu}} \int\frac{d^3q}%
{(2\pi)^3}B_{\alpha}({\bf q},\omega_{\mu},\omega_{\nu})L({\bf q},\omega_{\mu})\times 
\nonumber\\
\\
&\times &L({\bf q},\omega_{\mu}+\omega_{\nu})B_{\beta}({\bf q},\omega_{\mu},\omega_{\nu}),
\nonumber \eeq
The Green's functions block is defined as:
\beq \label{B}
B_{\alpha}({\bf q},\omega_{\mu},\omega_{\nu})&=&T\sum_{\omega_n}
\int\frac{d^3p}{(2\pi)^3}v_{\alpha}(p)C({\bf q},\omega_{n+\nu},\omega_{\mu-n})%
C({\bf q},\omega_n,\omega_{\mu-n})\times \nonumber\\
\\
&\times &G({\bf p},\omega_{n+\nu})G({\bf p},\omega_n)G({\bf q}-{\bf p},\omega_{\mu-n}). \nonumber
\eeq

In the vicinity of $T_{\rm c}$, for frequencies $\omega \ll T$, the leading 
singular contribution to the response operator $Q_{\alpha\beta}^{AL\,(R)}$ 
arises from the fluctuation propagators in (\ref{QAL}) rather than from the 
frequency dependences of the $B_\alpha$ blocks, so it suffices to neglect its 
frequency dependences \cite{AV80}. This approximation leads to 
\beq \label{approx} 
B_\alpha({\bf q},0,0)&=& -2\rho \frac{\eta}{v_{\rm F}^2}\frac{\partial}{\partial q_
\alpha}<[\xi({\bf p})-\xi({\bf q}-{\bf p})]^2>=\nonumber \\
\\
&=& -2\rho \frac{\eta}{v_{\rm F}^2}\cases{ sJ^2sin(q_zs) &for $\alpha=z$ \cr
v_{\rm F}^2q_{\alpha} & for $\alpha=x,y $}\nonumber \eeq

Using these expressions in (\ref{QAL}) followed by analitical continuation of 
the external Matsubara frequencies to the imaginary axis and integration over 
momenta, one can find the explicit expression for the imaginary part of the 
retarded electromagnetic response operator \cite{note2} for real frequencies
$\omega \ll T$:
\beq \label{QALperp} &&{\rm Im} \left[Q^{AL (R)}_{\perp}(\omega)\right]=\frac{e^2T}{4\pi%
s }\left(\frac{s^2}{\eta }\right)\left(\frac{16T_{\rm c}}{\pi \omega} \right) 
{\rm Re}\Bigg\{ \left(\frac{\pi \omega }{16T_{\rm c}}\right)^2 -\left(\epsilon %
-\frac{i\pi \omega }{16T_{\rm c}}+\frac r2\right)\times \nonumber \\
\\
&\times &\left[\Delta D_2\left(\epsilon -\frac{i\pi \omega }{16T_{\rm c}}\right)-%
\left(\frac{r}{2} \right)^2\Delta D_1\left(\epsilon -\frac{i\pi \omega }{16T_{\rm c}}
\right)\right] \Bigg\} \nonumber \eeq
\beq \label{QALparal}
&&{\rm Im} \left[Q^{AL (R)}_{\parallel}(\omega)\right]=\frac{2e^2T}{\pi s}{\rm Im} %
\Bigg\{\left[1+i\left(\frac{16T_{\rm c}}{\pi\omega}\right)\left(\epsilon+\frac r2%
\right)\right]\times\nonumber\\
\\
&\times &\left[\Delta D_1\left(\epsilon -\frac{i\pi\omega}{16T_{\rm c}}\right)
\right]+i\left(\frac{16T_{\rm c}}{\pi\omega}\right)\left[\Delta D_2\left(\epsilon %
-\frac {i\pi\omega}{16T_{\rm c}}\right)\right]\Bigg\}\nonumber \eeq
where $D_1(z)=2\ln\left[\sqrt{z}+\sqrt{\left(z+r\right)}\right]$, 
 $D_2(z)=-\sqrt{z(z+r)}$, $\Delta D_1\left(z\right)=D_1\left(z\right)
-D_1\left(\epsilon \right)$, $\Delta D_2\left(z\right)=D_2\left(z\right)-%
D_2\left(\epsilon \right)$ and $r=4\eta J^2/v_{\rm F}^2$.
The value $r\sim \xi_{\rm c}^2(0)/s^2$ is the usual anisotropy parameter \cite{Abrah} 
characterizing the dimensional crossover from the $2D$ to the $3D$ regime in 
the thermodynamic fluctuation behavior at $T_{\rm c}$ (except for $\sigma_{\perp}$, 
for which the crossover is from $0D$ to $3D$ at $T_{\rm c}$). The expressions 
presented above solve the problem of the frequency dependence of 
the paraconductivity tensor in the general form for $\epsilon \ll 1$ 
and $\omega \lesssim T$ for an arbitrary relation between 
$\epsilon, r$ and $\omega$, but they are too cumbersome.

Let us concentrate on the most interesting case for the HTS analysis of $2D$
fluctuations where $\xi_{\rm c}(T)\ll s$ ($r\ll \epsilon$) and 
$\sigma^{\rm AL}_{\perp}$
turns out to be suppressed by the necessity of the independent tunneling
of each electron participating in the fluctuation pairing from one $CuO_2$
layer to the neighbour one \cite{ILVY,Grenoble}.
The approximation $r\ll \epsilon $ simplifies considerably the expressions
(\ref{QALperp}) and (\ref{QALparal}) \cite{note3} and it keeps, in the same 
time, their validity up to frequencies comparable to $T_{\rm c}$:
\beq \label{perp} \sigma^{AL (2D)}_{\perp }(\epsilon ,\omega ) & = &
\frac{e^2s}{64\eta }\left( \frac{r}{2\epsilon }\right) ^2\displaystyle{ 
\frac 1{\tilde \omega^2}\ln \left( 1+\tilde \omega ^2\right) }= \nonumber\\
\\
& = & \sigma^{AL (2D)}_{\perp}(\epsilon ,0)%
\cases{ \displaystyle{ 1-\frac{\tilde \omega ^2}{2}} & for $\tilde \omega \ll 1$
\cr \displaystyle{ \frac{2}{\tilde \omega ^2}\ln {\tilde \omega}} & for %
$\tilde \omega \gg 1$} \nonumber \eeq and
\beq \label{abplane}\sigma ^{AL (2D)}_{\parallel}(\epsilon ,\omega )&=&          
\frac{e^2}{16s}\frac{1}{\epsilon }\left \{\frac{2}{\tilde \omega}\arctan 
\tilde \omega-\frac{1}{\tilde \omega ^2} \ln (1+\tilde \omega ^2) \right \}=%
\nonumber\\
\\
&=&\sigma ^{AL (2D)}_{\parallel }(\epsilon ,0)\cases{\displaystyle{%
1-\frac{\tilde \omega ^2}{6}} & for $\tilde \omega \ll 1$ \cr
\displaystyle{ \frac{\pi }{\tilde \omega }} & for $\tilde \omega \gg 1$}
\nonumber \eeq
where $\tilde \omega =\displaystyle{\frac{\pi \omega}{16(T-T_{\rm c})}}$.

Let us mention two facts following from the expressions obtained. Firstly, 
the paraconductivity begins to decrease rapidly with the increase of frequency 
already for $\omega \gtrsim T-T_{\rm c}$ (the critical exponents of this 
power decrease coincide with those ones in $\epsilon$-dependence of 
{\bf dc}-conductivity tensor components: $\nu_{\parallel} =1$ ($2D$ 
fluctuations) and $\nu_{\perp} = 2$ ($0D$ fluctuations)). Secondly the 
assumption to neglect the $\omega$-dependence of the Green's functions blocks 
evidently breaks down at frequencies $\omega \gtrsim T$ and this 
dependence has the only effect to accelerate the decrease. 

\section{Density of States Contribution}

As far as concern the DOS contribution to the electromagnetic
response operator tensor the four main diagrams for it (other two of this 
kind are negligible in the case under consideration \cite{BDKLV}) are 
presented in Fig. \ref{FedericiPRB.1} (2-5).
The general expression for the DOS contribution to 
$Q_{\alpha \beta }(\omega)$ from diagram 2 is:
\beq \label{QDOS}
Q_{\alpha \beta }^{DOS(2)}(\omega_{\nu})&=&2e^2T\sum_{\omega_{\mu}}
\int\frac{d^3q}{(2\pi)^3}L({\bf q},\omega_{\mu}) T\sum_{\omega_n}\int
\frac{d^3p}{(2\pi)^3}v_{\alpha }({\bf p})v_{\beta}({\bf p})\times \nonumber \\ 
\\
&\times &C^2({\bf q},\omega_n,\omega_{\mu-n})G^2({\bf p},\omega_n)G({\bf q}-
{\bf p},\omega_{\mu-n})G({\bf p},\omega_{n+\nu}),  \nonumber \eeq
The external frequency $\omega_{\nu}$ enters the expression (\ref{QDOS})
only by the Green's function $G({\bf p},\omega_{n+\nu})$ and it is not 
involved in ${\bf q}$ integration. So, near $T_{\rm c}$, even in the case of an 
arbitrary external frequency,  we can choose the propagator frequency 
$\omega_{\mu }=0$. In the same way  it can be treated the contribution from 
diagram 3 of Fig. \ref{FedericiPRB.1}.

The diagrams 2,4 and 3,5 of Fig. \ref{FedericiPRB.1}
are topologically equivalent and this fact would let think
that they give equal contributions to $\sigma(\omega)$.
Nevertheless the thoroughful analysis of the analitical 
continuation over the external frequency shows \cite{note4} 
that their contributions differ slightly and for the total DOS contribution 
to conductivity one can find:
\beq 
\label{ReDOSzzxx} 
{\rm Re}\left(\begin{array}{c} \sigma_{\perp}^{\rm DOS} 
(\omega)\\ \sigma_{\parallel}^{\rm DOS}(\omega) \end{array} \right) =%
-\displaystyle{\frac{e^2}{2\pi s}}\left( \begin{array}{c}\displaystyle{
\frac{s^2J^2}{v_{\rm F}^2}}\\ 1 \end{array}\right)\ln \left[\frac{2}
{\sqrt{\epsilon +r}+\sqrt{\epsilon}}\right]\kappa \left( \omega, T, \tau 
\right ), \eeq where
\beq 
\label{kappa} &&\kappa \left( \omega, T, \tau \right )=\displaystyle{
\frac{Tv_{\rm F}^2}{\eta}}\frac{1}{(\tau ^{-2}+\omega ^2)^2}\left\{\frac{4}{\tau }
\left[\psi \left(\frac12\right)-{\rm Re}\psi \left( \frac12 -\frac{i\omega}
{2\pi T}\right)\right]+
\frac{\tau^{-2}+\omega ^2}{4\pi T\tau }\frac{1}{\omega}
{\rm Im}\psi'\left(\frac12-\frac{i\omega}{2\pi T}\right)+%
\nonumber \right. \\
\\
 &+& \left. 
(\tau ^{-2}-\omega ^2)\frac1{\omega}\left[{\rm Im}\psi\left(\frac12-\frac{i\omega}
{2\pi T}\right)-2\,{\rm Im}\psi \left(\frac12-%
\frac{i\omega}{4\pi T}+\frac{1}{4\pi T\tau }\right)\right] \right \}. \nonumber \eeq
Let us stress that, in contrast to (\ref{QALperp}) and (\ref{QALparal}), this 
result has been found with the only assumption $\epsilon \ll 1$, so it is 
valid for any frequency, any impurity concentration and any dimensionality of 
the fluctuation behavior. The function $\kappa \left( \omega, T, \tau \right%
)$ can be easily used to fit experimental data. Nevertheless we present the 
asymptotics of the expression (\ref{kappa}) for clean and dirty cases.
In the dirty case
\beq \label{dirty} \kappa _{\rm d}\left( \omega ,T\ll \tau ^{-1}\right)= 
\displaystyle{\frac{Tv_{\rm F}^2}{2\eta}}\cases{\displaystyle{\frac{\tau }
{(2\pi T)^2}}\left |\psi''\left(\displaystyle{\frac12}\right)\right | & for 
$\omega \ll T\ll \tau^{-1}$ \cr \displaystyle{\frac{\tau}{\omega ^2}} & for 
$T\ll \omega \ll \tau ^{-1}$ \cr -\displaystyle{\frac{\pi}{\omega^3}} & 
for $T\ll \tau ^{-1}\ll \omega $} \eeq and in the clean case
\beq \label{clean}
\kappa _{\rm cl}\left( \omega ,T\gg \tau ^{-1}\right)= 
\displaystyle{\frac{Tv_{\rm F}^2}{2\eta}}\cases{\displaystyle{\frac{\pi \tau ^2}{4T}} 
& for $\omega \ll \tau ^{-1}\ll T$ \cr -\displaystyle{\frac{\pi }{4\omega ^2T}}
& for $\tau^{-1}\ll \omega \ll T$ \cr -\displaystyle{\frac{\pi}{\omega^3}}  
& for $\tau ^{-1}\ll T\ll \omega $} 
\eeq

\section{Maki-Thompson Contribution}

The total contribution of the MT-like diagrams to $\sigma_{\alpha \beta }%
(\omega)$ has been analysed in \cite{BDKLV} in the case of zero frequency and 
the frequency dependence of $\sigma _{\parallel}(\omega)$ has been studied in 
\cite{AV80}. In \cite{BDKLV} it was shown that actually the {\it regular} part 
of MT diagram can almost always be omitted. So we will not discuss it in 
this paper and we will concentrate on the {\it anomalous} MT contribution
(Fig. \ref{FedericiPRB.1} diagram 6).

In \cite{AV80} it was demonstrated that in the case of quasi two-dimensional
electron motion (\ref{corrug}) there is no formal necessity to introduce
the pairbreaking time $\tau _{\varphi}$ because the Maki-Thompson logarithmic 
divergency is automatically cut off due to the possible interlayer hopping.
Nevertheless, all evidences show that the intrinsic pairbreaking in HTS
is strong (at least one of its sources may be identified as thermal phonons) 
and an estimation of the appropriate 
$\tau_{\varphi}\sim 2\div 5\cdot 10^{-13}s$ is only 
several times larger than $T_{\rm c}^{-1}$. So we have to speak actually about 
the overdamped regime for Maki-Thompson contribution and it doesn't manifest 
itself noticeably in the $\epsilon $ dependence of conductivity \cite{BDKLV}
(the major part of experimental results is explained in terms of AL or AL 
and DOS contributions). Anyway we are interested in MT contribution 
because of its frequency dependence which evidently determines another 
characteristic scale in addition to the previous three 
($T-T_{\rm c}$, $T$, $\tau^{-1}$) we have introduced: $\omega _{\rm MT}\sim \tau %
_{\varphi }^{-1}$. 

We start, as usual, from the general expression of the 
{\it anomalous} MT contribution to the electromagnetic operator tensor 
\cite{BDKLV}:
\beq Q^{\rm MT}_{\alpha\beta}(\omega_{\nu})=2e^2T\sum_{\omega_{\mu}}%
\int\frac{d^3q}{(2\pi)^3}L({\bf q},\omega_{\mu})I_{\alpha\beta}
({\bf q},\omega_{\mu},\omega_{\nu}),\label{QMT} \eeq where
\beq \label{I}
I_{\alpha\beta}({\bf q},\omega_{\mu},\omega_{\nu})&=&T\sum_{\omega_n}\int
\frac{d^3p}{(2\pi)^3}v_{\alpha}({\bf p})v_{\beta}({\bf q}-{\bf p})C({\bf q},
\omega_{n+\nu},\omega_{\mu-n-\nu})C({\bf q},\omega_n,\omega_{\mu-n})\times\nonumber \\ 
\\
&\times &G({\bf p},\omega_{n+\nu})G({\bf p},\omega_n)G({\bf q}-{\bf p},%
\omega_{\mu-n-\nu})G({\bf q}-{\bf p},\omega_{\mu-n}).\nonumber \eeq
After the integration over ${\bf p}$ momentum and the summation over $\omega_n$ 
in the range $\omega_n \in [-\omega_{\nu },0[$ ({\it anomalous} part), 
one can find:
\beq \label{MTint} 
&&\left( \begin{array}{c} Q_{\perp}^{\rm MT(an)}(\omega _{\nu}) 
\\ Q_{\parallel}^{\rm MT(an)}(\omega _{\nu}) \end{array} \right)=e^2T\tau
\left[\psi\left(\frac12+\frac{\omega _{\nu}}{2\pi T}\right)-%
\psi \left(\frac12\right)\right]\times \nonumber \\
\\
&\times &\int \frac{d^3q}{(2\pi)^3}\left(\begin{array}{c}J^2s^2
\cos {q_{\perp}s}\\ v_{\rm F}^2 \end{array} \right) \frac{1}{\left(\omega_ {\nu }+%
\tau _{\varphi}^{-1}+D\hat Q^2 \right) \left(\epsilon +\eta q^2+r
\sin ^2(q_{\perp}s/2) \right) }\nonumber 
\eeq

At this stage of calculations we artificially introduce the phase-breaking 
time in the ``Cooperon" vertices. Carrying out the integration and 
separating the real and  the imaginary parts we have:
\beq \label{sigmaMT} &&{\rm Re}\left( \begin{array}{c} \sigma_{\perp}^{\rm MT(an)}
(\omega )\\
\sigma_{\parallel}^{\rm MT(an)}(\omega ) \end{array}\right)=
\frac{e^2}{2\pi s}\left(\begin{array}{c}\displaystyle{ \frac{s^2}{2\eta }}\\
1\end{array} \right)\frac{T}{\omega }{\rm Im}\left \{ \displaystyle{
\frac{ \psi \left(
\frac12-\frac{i\omega}{2\pi T}\right)-\psi \left( \frac12 \right)}
{\frac{i\pi \omega}{8T_{\rm c}}+\epsilon -\gamma}}
\times \nonumber \right. \\   
\\
&\times & \left.\left( \begin{array}{c}
-\Delta D_2\left(-\frac{i\pi \omega}{8T_{\rm c}}+\gamma \right)
+\left(\frac{i\pi \omega}{8T_{\rm c}}+\epsilon -\gamma \right) \\
\Delta D_1\left( %
-\frac{i\pi \omega}{8T_{\rm c}}+\gamma \right)
\end{array} 
\right) \right \} \nonumber 
\eeq
where 
$\gamma =\displaystyle{\frac{\pi }{8T_{\rm c}\tau_{\varphi}}}$.
In the case of two-dimensional overdamped regime 
($r\ll \epsilon \lesssim \gamma $) the expression (\ref{sigmaMT})
gives the following limits:
\beq \label{MTzzlim}\sigma_{\perp }^{MT(an)(2D)}(\omega)=\frac{e^2s}{2^7\eta}
\frac{r^2}{\gamma \epsilon }\cases{1 & for $\omega \ll \tau_{\varphi}^{-1}$
\cr \displaystyle{\left(\frac{8T_{\rm c}\gamma }
{\pi \omega }\right)^2} & for $\omega \gg \tau_{\varphi}^{-1}$}
\eeq
\beq \label{MTxxlim}
\sigma_{\parallel }^{MT(an)(2D)}(\omega)=\frac{e^2}{8s}
\cases{\displaystyle{\frac{1}{\gamma }}\ln 
\left(\displaystyle{\frac{\gamma}{\epsilon}}\right) & for
$\omega \ll \tau_{\varphi}^{-1}$ \cr 
\displaystyle{\frac{4T_{\rm c}}{\omega}} & for 
$\omega \gg \tau_{\varphi}^{-1}$}
\eeq

Let us remind that the expression (\ref{sigmaMT}) has been obtained without 
any limitation on frequency. Nevertheless we have done the assumption 
$\tau D\hat Q^2\ll 1$ through all the paper and it turns out that, while this 
condition doesn't restrict our results for the AL and the DOS contributions 
over the full range of frequency, temperature and impurity concentration, 
this is not so for the MT contribution. In fact, as it was shown in 
\cite{RanVar}, in the ultra-clean (or non-local) limit, when 
$T\tau >1/\sqrt \epsilon $, the assumption $\tau D\hat Q^2\ll 1$ is violated 
for MT contribution and the results obtained are not valid there. 
Nevertheless one can see that this 
non-local situation can be realized in the clean case ($T\tau \gg1$) only and 
for temperatures in the range $1/(T\tau )^2\ll \epsilon \ll 1$. We 
suppose $T\tau \sim 1$, as it is in the case of HTS, so we skip the 
discussion about the non-local limit.

\section{Discussion}

Let us start from the analysis of each fluctuation contribution separately and 
then we will discuss  their interplay in ${\rm Re}\left[\sigma_{\perp}(\omega)%
\right]$. 
Because of the large number of parameters entering the expressions we restrict 
our consideration to the $c$-axis component of the conductivity tensor in the 
region of $2D$ fluctuations (above Lawrence-Doniach crossover temperature). 
The in-plane component will be overviewed in the end of this section.

The AL contribution describes the fluctuation condensate response to the 
electromagnetic field applied. The component of the current associated with 
it can be treated as the precursor phenomenon of the screening currents
in the superconducting phase. Above $T_{\rm c}$ the virtual Cooper pairs binding 
energy gives rise to a pseudogap of the order of $T- T_{\rm c}$, so it is not 
surprising that at higher frequencies the AL contribution decreases with 
the further increase of $\omega$. 
Actually $\omega^{\rm AL}\sim T-T_{\rm c}$ is the only relevant 
scale for $\sigma^{\rm AL}$: its frequency dependence doesn't contain 
$T, \tau_\phi$ and $ \tau$. The independence from the latter manifests 
mathematically the fact that elastic impurities do not represent obstacles 
for the motion of Cooper pairs. The interaction of the electromagnetic 
wave with the fluctuation Cooper pairs resembles, in some way, the anomalous 
skin-effect where its reflection is determined by the interaction with the 
free electron system.

Another effect related with the formation of the fluctuation Cooper pairs,
but on the self-intersecting trajectories (like the weak localization 
correction), is described by the MT {\it anomalous} contribution.
Being the contribution related with the Cooper pairs electric charge transfer
it doesn't depend on the elastic scattering time but it turns out to be 
extremely sensitive to the phase-breaking mechanisms. So two  characteristic
scales turn out to be relevant in the frequency dependence of that one:
$T-T_{\rm c}$ and $ \tau_\phi^{-1}$. In the case of HTS, where $\tau_\phi^{-1}$ 
has to be estimated as at least $0.1 T_{\rm c}$ for temperatures up to 
$5\div 10\: K$
above $T_{\rm c}$, the MT contribution is overdamped, it is determined by the value
of $ \tau_\phi$ and it almost does not depend on temperature.

The density of states fluctuation renormalization gives quite different 
contribution to ${\rm Re}\left[\sigma(\omega)\right]$ with respect to those above. 
Physically it is related with the decrease of the one electron density due to 
the involvement of some number of electrons in the fluctuation Cooper pairing. 
At low frequencies ($\omega \ll \tau^{-1}$) the lack of electron states on the 
Fermi level leads to an opposite effect in comparison with AL and MT 
contributions: ${\rm Re}\left[\sigma^{\rm DOS}(\omega)\right]$ turns out to be negative 
and this means the increase of the surface impedance, or, in other words, 
the decrease of reflectance. Nevertheless, the electromagnetic field applied 
affects the electron distribution and at high frequencies 
$\omega \sim \tau^{-1} $ the DOS contribution changes its sign. 
It is interesting that DOS contribution, as one-electron effect, depends on 
the impurity  scattering similiarly to the normal Drude conductivity. 
The decrease of ${\rm Re}\left[\sigma^{\rm DOS}(\omega)\right]$ starts already 
at frequencies $\omega \sim \min\{T,\tau^{-1}\}$ which for HTS are much 
higher than $T-T_{\rm c}$ and $\tau_{\varphi}^{-1}$.

The scenario of ${\rm Re}[\sigma_{\perp}^{\rm tot}]$ $\omega $-dependence with 
the most natural choice of parameters ($r \ll \epsilon \lesssim 
\tau_{\varphi}^{-1}\ll \min\{T,\tau^{-1}\}$)
is presented in Fig. \ref{FedericiPRB.2}. The positive AL and MT effects, 
in their $\omega$-dependence, are well pronounced at low frequencies 
on the background of the DOS contribution which remains in this region
 a negative constant. Then at $\omega \sim T-T_{\rm c}$ the former decays  
and the ${\rm Re} \sigma_{\perp}$  remains negative up to
$ \omega \sim \min\{T,\tau ^{-1}\}$. The DOS 
correction changes its sign at $\omega \sim \tau^{-1}$ and then it rapidly 
decreases.
The following high frequency behavior is governed by the Drude law. So one 
can see that the characteristic pseudogap-like behavior in the frequency 
dependence of the 
optical conductivity, mentioned in the title of this paper, takes place in 
the range $\omega \in [T-T_{\rm c},\tau^{-1}]$. The depth of the window 
increases logarithmically with $\epsilon$ when $T$ tends to $T_{\rm c}$, as 
shown in Fig. \ref{FedericiPRB.3}.
In the case of $ab$-plane optical conductivity the two first
positive contributions are not suppressed by the interlayer transmittance,
they exceed considerably the negative DOS contribution in a wide range of 
frequencies and any pseudogap-like behavior is unlikely in $\sigma_%
{\parallel}^{\rm fl}(\omega)$: the reflectivity has to be of the metal kind. 
The comparison between (\ref{abplane}) and (\ref{ReDOSzzxx}) shows that the 
compensation of the two contributions could only take place at $\omega_0 \sim 
T/ \ln \epsilon$ which is out of the range of validity of the AL 
contribution. 

Let us compare now the results of our calculations with the experiments
available. The recent measurements \cite{Timusk1,Timusk2} of the $c$-axis 
reflectivity spectra, in the FIR region on $YBa_2Cu_4O_8$ single crystals, 
show the response of a poor metal with the additional contributions from $IR$
active phonon modes (which we do not discuss here). 
With the decrease of temperature 
the $c$-axis optical conductivity decreases showing a transition from a 
Drude-like to a pseudogap-like behavior at $\omega \sim 180\: cm^{-1}$. This 
gap grows deeper below $180\: K$ without any abrupt change at the 
superconducting transition temperature $T_{\rm c} = 80 K$.

Such experimentally observed behavior of the optical conductivity is in 
qualitative agreement with our results. Really the suppression of the density 
of states due to the superconducting fluctuations in the vicinity of $T_{\rm c}$ 
leads to the decrease of reflectivity in the range of frequencies up 
to $\omega \sim \tau^{-1}$. The magnitude of this depression slowly 
(logarithmically) increases with the decrease of temperature but, evidently, 
at the edge of the transition it reaches some saturation because of
the crossover to the $3D$ regime in fluctuations (where instead of 
$\ln (1/\epsilon)$ one has $\ln (1/r)-\sqrt{\epsilon}$, see 
(\ref{ReDOSzzxx})). So no singularity is expected in the value of the minimum 
even in the first order of perturbation theory. Below $T_{\rm c}$ the 
fluctuation behavior of $\langle\Delta_{\rm fl}^2\rangle$ is mostly symmetrical
to that one above $T_{\rm c}$ (see \cite{VD}) and, with the further decrease of
temperature, the fluctuation pseudogap minimum in the optical conductivity
smoothly transforms itself into the real superconducting gap, which  opens, 
in HTS, very sharply. Let us stress, that the independence of the pseudogap 
threshold from temperature appears naturally in our theory: it is determined
by $\omega_0\sim \tau ^{-1}$ (see (\ref{dirty}), (\ref{clean}) and 
Fig. \ref{FedericiPRB.4}). Comparing Fig. \ref{FedericiPRB.3} e Fig. \ref{FedericiPRB.4}, one can easy
see that the threshold doesn't move when the temperature changes but varies
when the inverse of the scattering rate changes.
As far as 
concern the numerical value of $\omega_0$, supposing $T_{\rm c}\tau=0.35$ 
(which is the value for the scattering rate of the sample used in the 
experiment under consideration \cite{Timusk2}, that is also in the 
experimental range of the inverse of the scattering rate 
$T\tau\approx 0.3\div 0.7$ \cite{Ginsberg,Calvani}), one can see that the 
pseudogap threshold is of the order of $200\: cm^{-1}$, in agreement with the 
experimental data, even from a quantitative point of view 
\cite{Timusk1,Timusk2}.

Our theory is, strictly speaking, valid only in the vicinity of the
critical temperature, where $\epsilon \ll 1$. Nevertheless the logarithmic
dependence on $\epsilon$ of the result obatained gives grounds to believe
that qualititatively the theory can be valid up to $\epsilon =\ln (T/T_{\rm c})\sim 1$,
so for temperatures up to $200\: K$ in the experiment discussed. So the
theory is again in agreement with the experimental value of temperature
$180\: K$ until which the pseudogap is observable.

In conclusion we have calculated the tensor of optical conductivity for
layered superconductors. The pseudogap-like minimum of its $c$-axis component 
in a wide range of frequencies for temperatures in the vicinity of 
$T_{\rm c}$ is found. 
Its origin is related with the fluctuation density of states
renormalization which can be treated as the opening of the fluctuation
pseudogap. These result are qualitatively, and in some aspects quantitatively, 
in agreement with the recent experiments on $YBa_2Cu_4O_8$ samples.
Further experiments, with more anisotropic samples like $BSSCO$ single 
crystals, would be useful, because the effect should be more prononced. 

\vskip 1cm

\centerline{\bf ACKNOWLEDGMENTS}

\vskip 0.6cm

We would like to thank P. Calvani, M. Capizzi, C. Di Castro, 
D. Livanov, M. Randeria and V. Tognetti for valuable discussions. This work 
was partially supported by NATO Collaborative Research Grant $\# CRG 941187$.

\newpage
\begin{figure}
\caption{We report the leading Feynman diagrams contributing to the optical 
conductivity. The wavy lines indicate the fluctuation propagators;
the thin solid lines with arrows stand for the impurity-averaged normal-state 
electron Green's functions; the shaded partial circles are the vertex 
corrections arising from impurities; the dashed curves with central crosses are
additional impurity renormalizations. Diagram 1 is the Aslamazov-Larkin 
contribution, diagrams 2-5 are the corrections from the density of states  
renormalization and diagram 6 is the Maki-Thompson diagram.}
\label{FedericiPRB.1}
\end{figure}
\newpage
\begin{figure}
\caption{The plot shows the dependence of the real part of conductivity,
normalized on the Drude normal conductivity, on $\omega/T$,
$\Re \left[\sigma^{\prime}(\omega)\right]={\rm Re}\left[\sigma (\omega)\right]%
/\sigma^{\rm n}$.
The dashed line refers to the $ab$-plane component of the conductivity tensor
whose Drude normal conductivity is 
$\sigma^{\rm n}_{\parallel}=\rho e^2\tau v_{\rm F}^2$.
The solid line refers to the $c$-axis component whose Drude normal 
conductivity is 
$\sigma^{\rm n}_{\perp}=\sigma^{\rm n}_{\parallel}J^2s^2/v_{\rm F}^2$.
In this plot we have put
 $T\tau=0.3, E_{\rm F}/T=50, r=0.01, \epsilon=0.04, T\tau_{\varphi}=4$.}
\label{FedericiPRB.2}
\end{figure}
\newpage
\begin{figure}
\caption{The behavior of the $c$-axis component of 
conductivity frequency dependence, for different values of temperature, 
is shown. The solid line refers to $\epsilon =0.04$; 
the dashed line refers to $\epsilon=0.06$; the dot-dashed line 
refers to $\epsilon =0.08$. $T\tau=0.2$ for all the curves.
The other parameters of this plot are the same used in 
Fig. \ref{FedericiPRB.2}.}
\label{FedericiPRB.3}
\end{figure}
\newpage
\begin{figure}
\caption{The plot shows the dependence of 
${\rm Re}\left[\sigma_{\perp}(\omega)/\sigma_{\perp}^{\rm n}\right]$ on 
$\omega/T$ for different values of $T\tau$. The solid line refers to
$T\tau =0.4$; the dot-dashed line refers to $T\tau =0.3$; the dashed line
refers to $T\tau =0.2$. The other parameters of this plot are the same used
in Fig. \ref{FedericiPRB.2}.}
\label{FedericiPRB.4}
\end{figure}

\begin{thebibliography}{99}
\bibitem{note2} It is necessary to mention that the direct calculation of the 
expression (\ref{QAL}) leads to the appearence of divergent expressions in the 
${\rm Re} Q$. Nevertheless, as it was shown in \cite{AV80}, the thorough 
summation of all diagrams from the first order of the perturbation theory for 
$Q$ before momentum integration leads to the exact cancellation of 
${\rm Re}\left[ Q^{\rm fl}(0)\right]$. 
Namely this fact justifies the possibility of the further 
calculation of different diagrammatic terms to $Q^{\rm fl}$ separately.
\bibitem{note3} The second expression coincides with that one obtained in 
\cite{AV80,Sch}.
\bibitem{note4} This fact had not been taken into account before and it
led to the lack of the factor $1/2$ in the dirty and clean
limit of the DOS contribution \cite{ILVY,BDKLV}.
We want to thank D. Livanov who attracted our attention to this circumstance.
\end{thebibliography}
\end{document}